\documentclass[apl,amsmath,amssymb,amnsfonts,twocolumn,a4paper]{revtex4-1}
\usepackage{etex}
\usepackage[T1]{fontenc}
\usepackage{geometry}
\usepackage{graphicx}
\usepackage{color}
\usepackage[usenames,dvipsnames]{pstricks}
\usepackage{epsfig}

\usepackage{setspace}
\usepackage{booktabs}
\usepackage[strict]{changepage}

\usepackage{todonotes}

\begin{document}

\title{Basic noise mechanisms of heat-assisted-magnetic recording} 

\author{Christoph Vogler}
\email{christoph.vogler@tuwien.ac.at}
\affiliation{Institute of Solid State Physics, TU Wien, Wiedner Hauptstrasse 8-10, 1040 Vienna, Austria}
\affiliation{Institute of Analysis and Scientific Computing, TU Wien, Wiedner Hauptstrasse 8-10, 1040 Vienna, Austria}

\author{Claas Abert}
\author{Florian Bruckner}
\author{Dieter Suess}
\affiliation{Christian Doppler Laboratory for Advanced Magnetic Sensing and Materials, Institute for Solid State Physics, TU Wien, Wiedner Hauptstrasse 8-10, 1040 Vienna, Austria}

\author{Dirk Praetorius}
\affiliation{Institute of Analysis and Scientific Computing, TU Wien, Wiedner Hauptstrasse 8-10, 1040 Vienna, Austria}

\date{\today}

\begin{abstract}
Heat-assisted magnetic recording (HAMR) is expected to be a key technology to significantly increase the areal storage density of magnetic recording devices. At high temperatures thermally induced noise becomes a major problem, which must be overcome in order to reliably write magnetic bits with narrow transitions. We propose an elementary model based on the effective recording time window (ERTW) to compute the switching probability of bits during HARM of bit-patterned media. With few assumptions this analytical model allows to gain deeper insights into the basic noise mechanisms like AC and DC noise. Finally, we discuss strategies to reduce noise and to increase the areal storage density of both bit-patterned as well as granular media.
\end{abstract}

\maketitle 

\section{Introduction}
\label{sec:intro}
Decreasing grain size and at the same time increasing magnetic anisotropy to maintain thermal stability yields increasing areal storage densities of magnetic storage devices. However, high magnetic anisotropy results in high coercivity, which is a problem for available magnetic write fields of recording heads. Heat-assisted magnetic recording (HAMR)~\cite{mee_proposed_1967,guisinger_thermomagnetic_1971,kobayashi_thermomagnetic_1984,rottmayer_heat-assisted_2006} locally reduces the coercive field beyond available write fields of bits in order to reliably switch their magnetization. Recording at high temperatures near or above the Curie temperature $T_{\mathrm{C}}$ of the involved grains is a source of thermal noise, and thus thermally induced errors. There are two main noise sources to be distinguished: (i) AC noise, which determines the width of transitions in granular media and the distance between neighboring bits in bit-patterned media. (ii) DC noise, which restricts the maximum switching probability of magnetic grains away from transitions in granular media and the overall maximum switching probability of bits in bit-patterned media. HAMR was first proposed almost 60 years ago. Nevertheless, there exists little knowledge about the detailed noise mechanisms. To increase the areal storage density such knowledge is essential to further improve HARM devices beyond the recently reached 1.4\,Tb/in$^2$~\cite{ju_high_2015} of a working prototype.

Zhu and Li~\cite{zhu2013understanding,zhu2015medium} demonstrated a clear correlation between noise power and the recording time window for FePt-L$_0$ thin film granular media. Calculating signal-to-noise ratios of granular media is computationally challenging. Due to the large number of free input parameters it is hard to find influences of single parameters on the produced noise. The investigation of bit-patterned media allows for detecting such connections and permits conclusions to be drawn about granular media. 

The structure of the paper is as follows: Section~\ref{sec:HAMR_techniques} introduces the used model of HAMR. Additionally the definition of the effective recording time window (ERTW) is given. In Sec.~\ref{sec:switchingProb} switching probabilities of recording grains based on the ERTW approach are computed and compared to direct calculations by means of the stochastic Landau-Lifshitz-Bloch equation. Basic noise mechanisms are also explained on the basis of simple examples. Section~\ref{sec:Conclusion} concludes the findings and gives advices to reduce noise in bit-patterned and granular media.

\section{HAMR techniques}
\label{sec:HAMR_techniques}
We assume a continuous laser spot with a Gaussian spatial distribution and a full width at half maximum (FWHM) of 20\,nm, which moves in down-track direction $x$ over the medium with a velocity $v_{\mathrm{h}}$ per:
\begin{equation}
\label{eq:gauss_profile_CLSR}
 T(x,t)=\left ( T_{\mathrm{peak}}-T_{\mathrm{min}} \right )e^{-\frac{\left (x-v_{\mathrm{h}}t   \right )^2}{2\sigma^2}}+T_{\mathrm{min}},
\end{equation}
with
\begin{equation}
 \sigma=\frac{\mathrm{FWHM}}{\sqrt{8\ln(2)}}.
\end{equation}
$T_{\mathrm{peak}}$ and $T_{\mathrm{min}}$ are the maximum and minimum temperatures of the heat pulse. According to Eq.~\ref{eq:gauss_profile_CLSR} the heat pulse has a Gaussian shape in time as well. 

An external magnetic field is applied to the recording grains in order to reverse their magnetization. The external field has a trapezoidal shape with a base duration in write direction of 1\,ns and a field rise and decay time of 0.1\,ns, respectively. The field is symmetric around $t=0$ and has an amplitude of 0.8\,T. At the beginning and the end of each simulation it points in the $+z$ direction, in between it points for 1\,ns in the $-z$ direction, which is the desired write direction.

\subsection{effective recording time window (ERTW)}
In this work we aim to investigate the influence of the effective recording time window (ERTW) on the switching probability of recording grains in bit-patterned media during HAMR. A significant correlation between ERTW and signal-to-noise ratio was already reported in Refs.~\cite{zhu2013understanding,zhu2015medium} for HAMR of granular media. We want to go a step further and show that the basic noise mechanisms of HAMR can be described almost entirely with the ERTW. 

The magnetization of high anisotropy magnetic materials, like FePt, cannot be reversed with state-of-the-art head fields at room temperature. At higher temperatures the coercivity decreases until it falls below the available magnetic write field. We refer to this temperature as freezing temperature $T_{\mathrm{f}}$ in the following. Hence, we define the recording temperature window $\tau_{\mathrm{rec}}$ as 
\begin{equation}
 \label{eq:recordingTempWin}
 \tau_{\mathrm{rec}}=t(T_{\mathrm{C}})-t(T_{\mathrm{f}}),
\end{equation}
where $t(T_{\mathrm{C}})$ and $t(T_{\mathrm{f}})$ denote the times when the applied temperature decreases below the Curie point and the freezing temperature, respectively. Here, just the transition from the paramagnetic to the ferromagnetic state is considered, because all information which is written during heating is lost after exceeding $T_{\mathrm{C}}$. $\tau_{\mathrm{rec}}$ strongly depends on the shape of the applied heat pulse and the head velocity.

\begin{figure}[!h]
\centering
\includegraphics{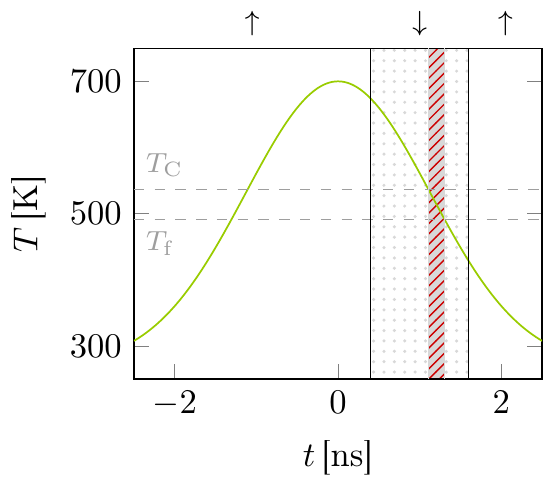}
\caption{\small (color online) Definition of the effective recording time window (ERTW). The green line shows the applied temperature pulse, the grey solid area denotes $\tau_{\mathrm{rec}}$ (see Eq.~\ref{eq:recordingTempWin}) and the grey dotted area illustrates the time span for which the external magnetic field is applied in write direction. ERTW$_\downarrow$ is defined as the overlap of the two latter and is marked with a red striped area.}
\label{fig:illustration}
\end{figure}

Finally, we define the ERTW as intersection of $\tau_{\mathrm{rec}}$ and the time period in which the external magnetic field points in $\phi$ direction, which is the $\uparrow$ or the $\downarrow$ direction in our analysis: 
\begin{equation}
 \label{eq:ERTW}
 \text{ERTW}_\phi=\tau_{\mathrm{rec}} \cap \left ( t_{\phi,\mathrm{final}}-t_{\phi,\mathrm{start}}\right).
\end{equation}
Here, $t_{\phi,\mathrm{start}}$ and $t_{\phi,\mathrm{final}}$ denote the start and the end time of the magnetic field in $\phi$ direction, respectively. For illustration purposes Fig.~\ref{fig:illustration} exemplarily illustrates the ERTW for a heat pulse with a peak temperature of 700\,K. The solid grey area displays $\tau_{\mathrm{rec}}$ and the grey dotted area shows the time span of the magnetic field pulse in write direction (the field directions are indicated by the arrows above the plot). The overlap denotes ERTW$_{\downarrow}$, which is highlighted with a red striped area.

\section{Switching Probabilities}
\label{sec:switchingProb}
\begin{figure*}[t]
\includegraphics{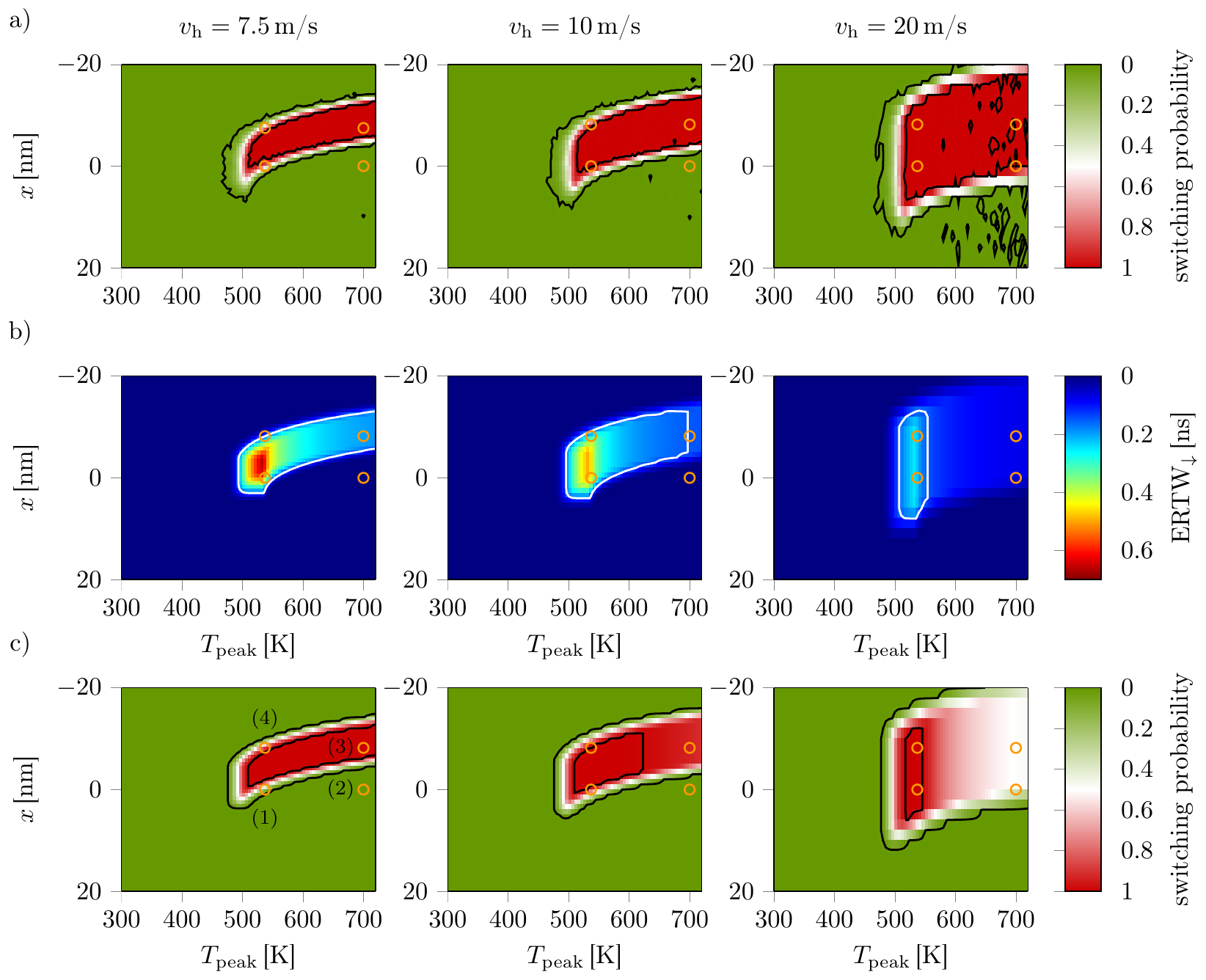}
  \caption{\small (color online) a) Switching probability phase diagrams of a hard magnetic recording grain (see Tab.~\ref{tab:prop}) under an applied field of 0.8\,T. Various head velocities are compared. The probabilities were calculated with a coarse-grained LLB model (Ref.~\cite{volger_llb}) and were taken from Ref.~\cite{vogler_areal_2015}. b) ERTW$_\downarrow$ for the same recording parameters calculated with Eqs.~\ref{eq:recordingTempWin} and \ref{eq:ERTW} and $T_{\mathrm{f}}=491.5$\,K. c) Switching probabilities computed per Eq.~\ref{eq:ERTW_prob} based on the ERTW. The contour lines in a) and c) mark the transition between areas with less than 1\,\% switching probability (light green) and areas with more than 99\,\% probability (dark red). The white contour lines in b) denote lines with $\text{ERTW}_\downarrow=\theta_{\mathrm{ERTW}}=0.15$\,ns.}
  \label{fig:clsr_vel_comp}	
\end{figure*}
In Ref.~\cite{vogler_areal_2015} the switching probabilities of a hard magnetic cylindrical recording grain with a diameter of 5\,nm and a height of 10\,nm, consisting of the material defined in Tab.~\ref{tab:prop}, were calculated with a coarse-grained Landau-Lifshitz-Bloch model for various peak temperatures $T_{\mathrm{peak}}$ and down-track positions $x$.
\begin{table}[h!]
  \centering
  \begin{tabular}{c c c c c}
    \toprule
    \toprule
      $K_1$\,[J/m$^3$] & $J_{\mathrm{S}}$\,[T] & $A_{\mathrm{ex}}$\,[pJ/m] & $T_{\mathrm{C}}$\,[K] & $\lambda$\\
    \midrule
      $6.6\times10^6$ & 1.43 & 21.58 & 536.94 & 0.1\\
    \bottomrule
    \bottomrule
  \end{tabular}
  \caption{\small Magnetic properties of the investigated hard magnetic recording grain.}
  \label{tab:prop}
\end{table}
In the following we demonstrate that the switching probability of a recording grain subject to an applied heat pulse and an additional external magnetic field is determined by the duration of the ERTW. The switching probabilities for 3 different head velocities displayed in Figure~\ref{fig:clsr_vel_comp}a are taken from Ref.~\cite{vogler_areal_2015}. The black solid lines mark the transition areas between 0\,\% and 99\,\% switching probability. The calculated ERTW$_\downarrow$ for the same recording parameters is illustrated in Fig.~\ref{fig:clsr_vel_comp}b. The obtain values vary in the range of $0.0-0.7$\,ns. As expected smaller head velocities yield larger ERTW$_\downarrow$ because of the larger $\tau_{\mathrm{rec}}$. For all considered head velocities ERTW$_\downarrow$ is the largest for peak temperatures around {$T_{\mathrm{C}}$. Both phenomena are a result of the smaller temporal thermal gradients. The white contour lines in Fig.~\ref{fig:clsr_vel_comp}b indicate ERTW$_\downarrow=0.15$\,ns. 

For the further computation we assume a linear dependence between ERTW and the bit's switching probability per:
\begin{equation}
\label{eq:ERTW_prob}
 p=\min \left (\frac{\mathrm{ERTW}_{\downarrow}}{\theta_{\mathrm{ERTW}}},1 \right ) \left [ 1-\min \left (\frac{\mathrm{ERTW}_{\uparrow}}{\theta_{\mathrm{ERTW}}},1 \right ) \right ].
\end{equation}
Here, we assume a threshold value of $\theta_{\mathrm{ERTW}}=0.15$\,ns, for which complete switching occurs. The choice of the threshold is motivated by the optimal recording time window between 0.1\,ns and 0.2\,ns in Ref.~\cite{zhu2013understanding} for FePt-L$_0$ thin film granular media. The first term of Eq.~\ref{eq:ERTW_prob} denotes the probability to write the heated recording bit in write direction, which is the $\downarrow$ direction. Since, the probability cannot exceed $p=1$ and an $\text{ERTW}$ of 0.15\,ns is considered as threshold value for complete switching of the bit, the minimum function is used. One has to take into account that a bit, which was previously written in $\downarrow$ direction can be overwritten after the reversal of the external field. Hence, the probability for this reversal is represented by the second term of Eq.~\ref{eq:ERTW_prob}. The equation actually describes the joint probability of aligning the recording bit in write direction and not revering it afterwards. The probabilities are just based on the according ERTW. 

\subsection{freezing temperature $T_{\mathrm{f}}$}
 \label{sec:freezingTemp}
To calculate the ERTW per Eq.~\ref{eq:ERTW} via Eq.~\ref{eq:recordingTempWin} the freezing temperature, for a maximum write field of 0.8\,T, must be estimated. We performed several hysteresis loop simulations with the introduced hard magnetic material (see Tab.~\ref{tab:prop}) under various temperatures, in order to extract the temperature dependence of the coercive field. In detail, we integrated the stochastic Landau-Lifshitz-Bloch equation~\cite{garanin_thermal_2004,chubykalo-fesenko_dynamic_2006,atxitia_micromagnetic_2007,kazantseva_towards_2008,chubykalo-fesenko_dynamic_2006,schieback_temperature_2009,bunce_laser-induced_2010,evans_stochastic_2012,mcdaniel_application_2012,greaves_magnetization_2012,mendil_resolving_2014} as described in Ref.~\cite{volger_llb} for hysteresis loops in the field range of 5\,T to -15\,T with a sweep rate of 100\,mT/ns. For each temperature value we computed 128 loops to obtain better statistics. Temperatures with a resulting coercivity lower than 0.8\,T were counted as possible write temperatures.
\begin{figure}[!h]
\centering
\includegraphics{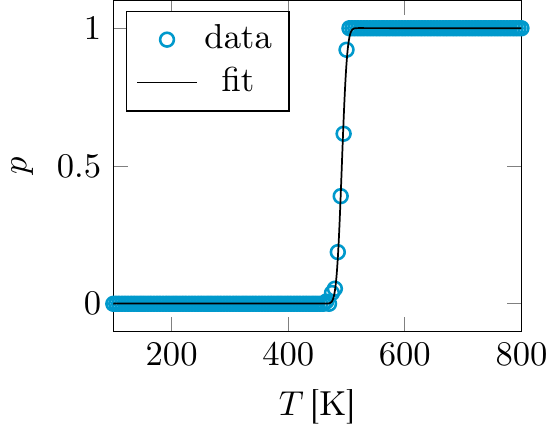}
\caption{\small (color online) Probability to obtain a coercive field of 0.8\,T or less, if a hysteresis loop in the field range of 5\,T to -15\,T with a sweep rate of 100\,mT/ns is simulated at various constant temperatures. A grain with the material parameters of Tab.~\ref{tab:prop} is assumed. The black solid line shows a fit with the cumulative distribution function of the normal distribution.}
\label{fig:freezing_field}
\end{figure}
The goal of this procedure was to determine the probability of lowering the coercive field of the grain below 0.8\,T depending on the applied temperature, which is displayed in Fig.~\ref{fig:freezing_field}. The data were fitted with the cumulative distribution function of the normal distribution to extract mean value and standard deviation of the freezing temperature to $491.5 \pm 7.1$\,K.\newline

With the known value and distribution of the freezing temperatures, ERTW$_\downarrow$ displayed in Fig.~\ref{fig:clsr_vel_comp}b was used to compute switching probabilities of the hard magnetic recording bit, with Eqs.~\ref{eq:recordingTempWin} and ~\ref{eq:ERTW}, as illustrated in Fig.~\ref{fig:clsr_vel_comp}c. The directly calculated switching probabilities of Fig.~\ref{fig:clsr_vel_comp}a and those obtained from the ERTW approach of Fig.~\ref{fig:clsr_vel_comp}c agree surprisingly well. Both, the shape of the core with more than 99\,\% switching probability and the transition area show the same behavior for all investigated head velocities. The ERTW approach is also capable of reproducing the DC noise at high peak temperatures which arises for head velocities of 10\,m/s and 20\,m/s. To acquire a deeper understanding of the underlying mechanisms we investigated 4 representative phase points, marked with orange circles in Fig.~\ref{fig:clsr_vel_comp}, in more detail. 

We start with a head velocity of 7.5\,m/s.
\begin{figure}
\includegraphics{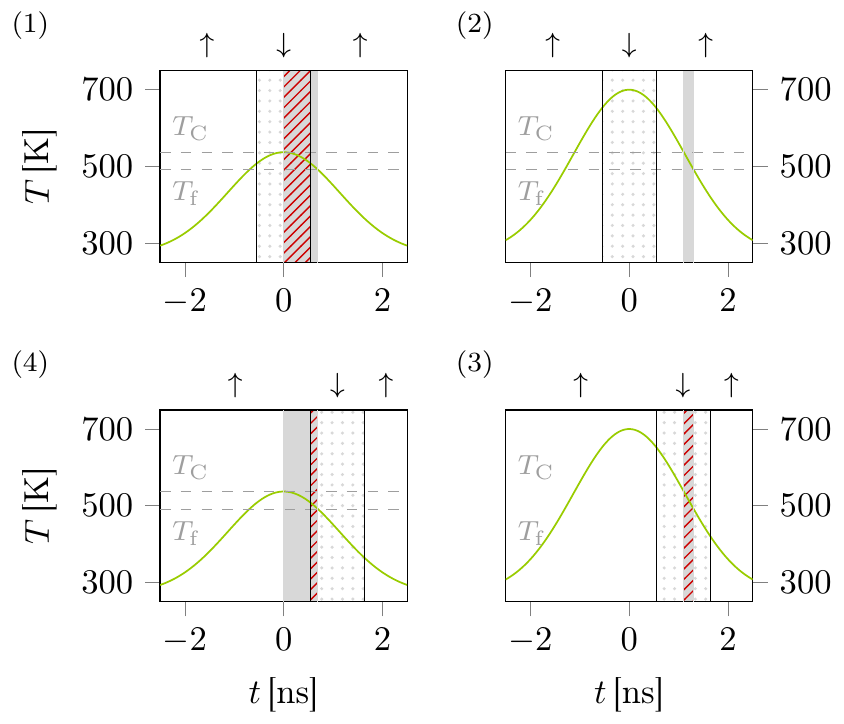}
    \caption{\small (color online) ERTW illustrations for the marked phase points $(1)-(4)$ in the first column of Fig.~\ref{fig:clsr_vel_comp}. A head velocity of $v_{\mathrm{h}}=7.5$\,m/s is assumed. (1) $T_{\mathrm{peak}}=T_{\mathrm{C}}$ and $x=0$\,nm (2) $T_{\mathrm{peak}}=700$\,K and $x=0$\,nm (3) $T_{\mathrm{peak}}=700$\,K and $x=-8.2$\,nm (4) $T_{\mathrm{peak}}=T_{\mathrm{C}}$ and $x=-8.2$\,nm. The plots are of the same type as shown in Fig.~\ref{fig:illustration}.}
  \label{fig:time_win_v7p5}	
\end{figure}
The arrangement of the plots in Fig.~\ref{fig:time_win_v7p5} corresponds to that of the 4 marked phase points in Fig.~\ref{fig:clsr_vel_comp}. All following plots are of the same type as introduced in Fig.~\ref{fig:illustration}. Figure~\ref{fig:time_win_v7p5}(2) illustrates a simulation with $T_{\mathrm{peak}}=700$\,K and $x=0$\,nm. During the whole time span $\tau_{\mathrm{rec}}$ (solid grey area) the external field points in $\uparrow$ direction. ERTW$_\downarrow$ is 0\,ns, and thus the probability for the bit's magnetization to end up in $\downarrow$ direction, for these recording parameters, is 0\,\%.

For a down-track position of $x=-8.2$\,nm the case is different. As shown in Fig.~\ref{fig:time_win_v7p5}(3) the external field points in write direction during the whole recording temperature window $\tau_{\mathrm{rec}}=0.2$\,ns. Hence, $\text{ERTW}_\downarrow=\tau_{\mathrm{rec}}$ is valid, yielding complete switching with a probability of more than 99\,\%. 

In the case of a peak temperature of $T_{\mathrm{peak}}=T_{\mathrm{C}}$ and a down-track position of $x=-8.2$\,nm, ERWT$_\downarrow$ equals $\theta_{\mathrm{ERTW}}=0.15$\,ns (see Fig.~\ref{fig:time_win_v7p5}(4)). Based on the threshold value one could expect the switching probability to be $p=1$,
but as Fig.~\ref{fig:clsr_vel_comp}c points out the phase point is located in the transition area. There are 2 reasons. First, and most important Fig.~\ref{fig:time_win_v7p5} just illustrates the situation for a fixed $T_{\mathrm{f}}$ of 491.5\,K. If the freezing temperature increases ERTW$_\downarrow$ decreases and falls below 0.15\,ns. Thus, a broadening of the transition area arises, which is a direct result of the fundamental distribution of the freezing temperature or the coercive field at high temperatures. Without $T_{\mathrm{f}}$ distribution the transition jitter would be vanishing. The second effect is not captured with the ERTW approach and just appears in the direct calculations shown in Fig.~\ref{fig:clsr_vel_comp}a. After the phase transition from the paramagnetic to the ferromagnetic state, below $T_{\mathrm{C}}$, the magnetic field points in $\uparrow$ direction. Before the field switches it aligns the magnetization of the bit antiparallel to the actual write direction (ERTW$_\uparrow=0.54$\,ns). Hence, after the field reversal more time is needed to switch the biased bit compared to the case where the magnetization gets aligned from the beginning of the ferromagnetic phase. 
Nevertheless, the ERTW jitter is one of the main sources of AC noise, which must be minimized to maximize the areal storage density of both granular and bit-patterned media.

In Fig.~\ref{fig:time_win_v7p5}(1) the phase point with $T_{\mathrm{peak}}=T_{\mathrm{C}}$ and $x=0$\,nm is displayed. Here, ERTW$_\downarrow=0.55$\,ns is more than three times larger than the threshold value. However, the switching probability is very low with $p\sim0.2$. The reason is that after ERTW$_\downarrow$ the temperature is sill below $T_{\mathrm{f}}$, and thus the reversed external field can try to overwrite the magnetization of the bit for ERTW$_\uparrow=0.14$\,ns until $\tau_{\mathrm{rec}}$ ends. Hence, the total probability of aligning the bit in write direction significantly lowers. Additionally $\tau_{\mathrm{rec}}$ jitters due to the distribution of $T_{\mathrm{f}}$, which further influences the probability. It has to be noted that only large ERTW$_\downarrow$ values do not guarantee high switching probabilities. It seems to be optimal to have $\text{ERTW}_\downarrow=\tau_{\mathrm{rec}}$. The window duration should be slightly above the threshold value for complete switching, and should be located in the middle of the external write pulse. Too large $\tau_{\mathrm{rec}}$ increases the probability to bias or to overwrite the magnetization of the recording bit. Similar observation can also be found in Ref.~\cite{zhu2015medium} regarding the relation between the duration of the recording time window and the signal-to-noise ratio of granular media.

\begin{figure}
\includegraphics{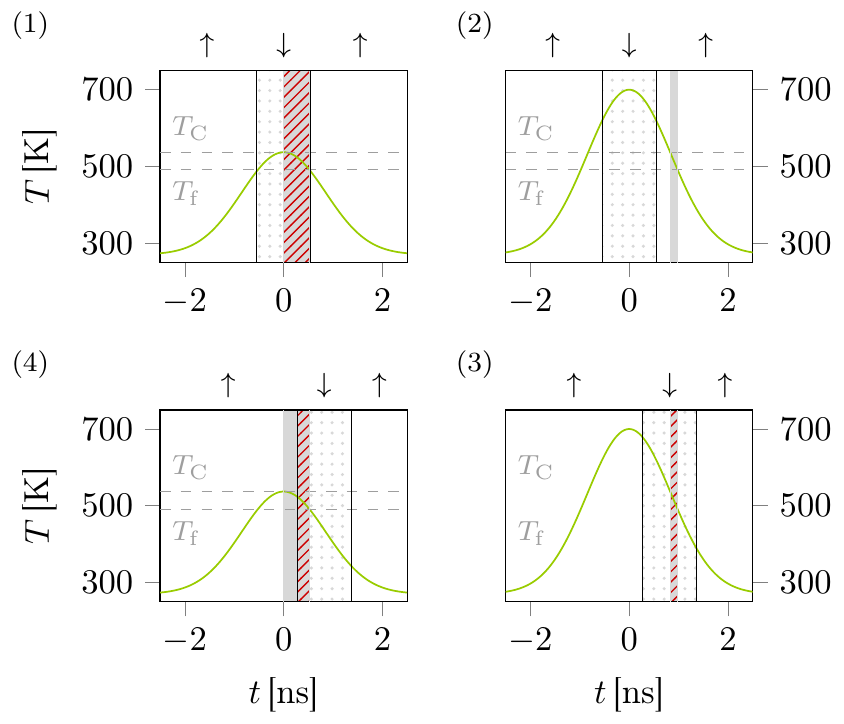}
    \caption{\small (color online) Same ERTW illustrations as displayed in Fig.~\ref{fig:time_win_v7p5} for a head velocity of $v_{\mathrm{h}}=10$\,m/s. The plots correspond to the marked phase points in the second column of Fig.~\ref{fig:clsr_vel_comp}.}
  \label{fig:time_win_v10}
\end{figure}

If a head velocity of $v_{\mathrm{h}}=10$\,m/s is used during recording a higher maximum thermal gradient in time arises at the bits. We analyze the same phase points as for $v_{\mathrm{h}}=7.5$\,m/s. For $T_{\mathrm{peak}}=700$\,K the plots look very similar to those with a lower head velocity. In the case of a down-track-position of $x=0$\,nm we obtain $\text{ERTW}_\downarrow=0$\,ns, yielding 0\,\% switching probability (see Fig.~\ref{fig:time_win_v10}(2)). Due to the higher thermal gradient $\tau_{\mathrm{rec}}$ and ERTW$_\downarrow$ decrease to 0.15\,ns for $x=-8.2$\,nm (see Fig.~\ref{fig:time_win_v10}(4)). The switching probability decreases below 95\,\% as a result of the ERTW jitter. The source of this DC noise is again the ERTW jitter, if the recording time window has a value near the threshold. 

In the case of a lower peak temperature ($T_{\mathrm{peak}}=T_{\mathrm{C}}$) no new physics can be observed. The large ERTW$_\downarrow=0.25$\,ns yields complete switching for $x=-8.2$\,nm based on the ERTW approach (see Fig.~\ref{fig:clsr_vel_comp}c). Here, a slight discrepancy occurs between the presented ERTW approach and the direct Landau-Lifshitz-Bloch simulations ((see Figure~\ref{fig:clsr_vel_comp}a)), where the phase point already shows AC noise. The reason is that the pre-alignment of the bit in $\uparrow$ direction during ERTW$_\uparrow=0.27$\,ns, before its magnetization can be reversed in $\downarrow$ direction, significantly increases the needed ERTW$_\downarrow$ for complete switching.
A large ERTW$_\downarrow$ of 0.52\,ns is obtained for a down-track position of $x=0$\,nm (see Fig.~\ref{fig:time_win_v10}(1)). Jitter of $\tau_{\mathrm{rec}}$ combined with overwriting issues are again source of the observed AC noise. 

\begin{figure}
\includegraphics{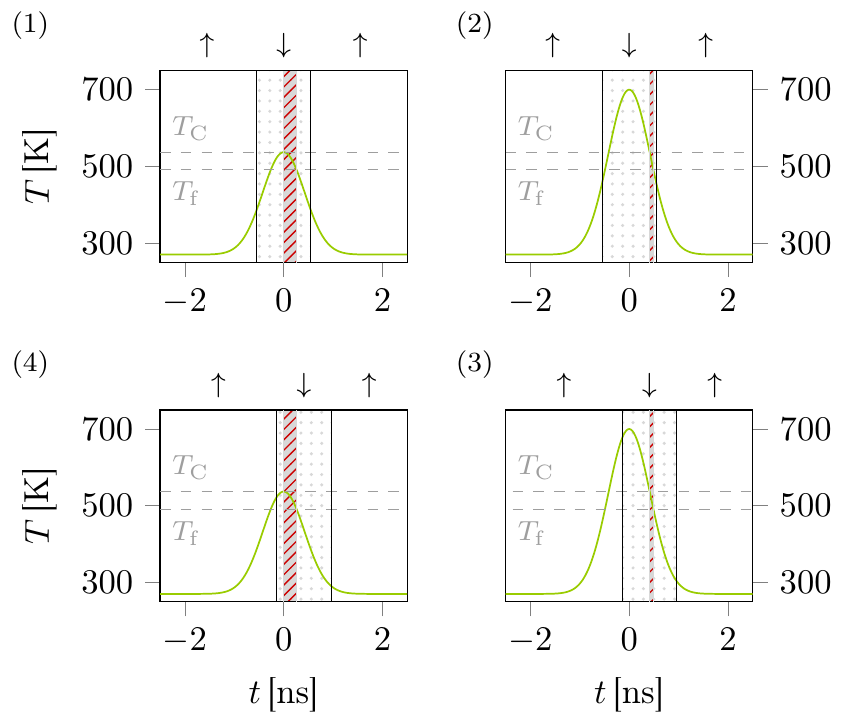}
    \caption{\small (color online) Same ERTW illustrations as displayed in Fig.~\ref{fig:time_win_v7p5} for a head velocity of $v_{\mathrm{h}}=20$\,m/s. The plots correspond to the marked phase points in the last column of Fig.~\ref{fig:clsr_vel_comp}.}
  \label{fig:time_win_v20}
\end{figure}
Finally, the phase points for $v_{\mathrm{h}}=20$\,nm are discussed on the basis of Fig.~\ref{fig:time_win_v20}. Because of the high thermal gradient $\tau_{\mathrm{rec}}$ is very narrow for $T_{\mathrm{peak}}=700$\,K. The resulting ERTW$_\downarrow=0.08$\,ns generates large DC noise for both investigated down-track positions $x$. For $T_{\mathrm{peak}}=T_{\mathrm{C}}$ we obtain $\text{ERTW}_\downarrow=0.26$\,ns. Since ERTW$_\uparrow=0$\,ns and ERTW$_\downarrow$ is large enough that the resulting switching probability is not influenced by ERTW jitter no AC noise is observed in these phase points and complete switching occurs.

It has to be mentioned that at high peak temperatures a mismatch in the DC noise level of the switching probabilities between direct simulations (Fig.~\ref{fig:clsr_vel_comp}a) and calculations based on the ERTW (Fig.~\ref{fig:clsr_vel_comp}c) is observed. This mismatch arises from the fact that even without external field the switching probability is at least 50\,\%, because after heating the material above its Curie temperature it is equally probable to end up in the $\uparrow$ or the $\downarrow$ state. Once an external field is applied the probability can only increase. Hence, the probabilities above $T_{\mathrm{C}}$ need a rescaling. Nevertheless, the simple model qualitatively agrees very well for all peak temperatures and down-track positions with the direct simulations via the integration of the stochastic Landau-Lifshitz-Bloch equation.

\section{Conclusion}
\label{sec:Conclusion}
In this work we analyzed the influence of the effective recording time window ERTW on the switching probability of a recording bit during heat-assisted magnetic recording (HMAR) of bit-patterned media. The ERTW was defined as the intersection between the time span $\tau_{\mathrm{rec}}$ during cooling of a bit in which the reduced coercive field is lower than the magnetic write field, and the time span during which the field points in write direction. We introduced a threshold value of ERTW in order to compute the switching probabilities of recording grains for various parameters. The probabilities based on the ERTW approach qualitatively agreed very well with those calculated via the integration of the stochastic Landau-Lifshitz Bloch equation.

Further, the ERTW approach allowed us to gain deeper insights into the basic noise mechanisms during HAMR. The detected jitter of $\tau_{\mathrm{rec}}$ plays a major role for both AC and DC noise. The jitter is a result of the thermally induced distribution of the coercive field at high temperatures. In combination with bias or overwriting effects the observed $\tau_{\mathrm{rec}}$ distribution yields incomplete switching, and thus AC noise at the transitions between bits. At high peak temperatures high thermal gradients lower the ERTW beyond its threshold value, yielding incomplete switching and DC noise. This effect is deteriorated by ERTW jitter. Hence, both noise mechanisms can be explained by means of the length of the ERTW.

Based on these findings, we propose several possibilities to influence noise during HAMR. Increasing the magnetic write field increases the ERTW, and thus reduces DC noise. In contrast AC noise increases because bias and overwriting effects gain in importance for increasing write fields. Similar limits were discussed in Ref.~\cite{suess_fundamental_2015}, where AC noise was referred to transition jitter and DC noise to written-in errors. As a consequence one must find a balance between minimum AC noise, which allows to write narrow transitions and minimum DC noise, which allows for complete switching of the involved recording bits. This holds for both granular and bit-patterned media. 

Zhu and Li~\cite{zhu2015medium} proposed to relax the anisotropy temperature gradient of recording grains to ensure high signal-to-noise ratios (SNR) for low write fields. According to Ref.~\cite{zhu2015medium} this is valid for devices with low to intermediate areal densities, where the SNR is grain pitch limited. Corresponding to our analysis this enlarges the ERTW and reduces DC noise for a given write field. As long as the SNR is grain pitch limited no broadening of the transition due to AC noise occurs for granular media. In terms of bit-patterned media the involved AC noise increase enlarges the transitions. As long as the bit distances are lager than transition jitter, AC noise has no influence on the areal storage density. Hence, a smaller anisotropy temperature gradient improves the bit error rate. As shown in Ref.~\cite{vogler_areal_2015} AC noise becomes an issue just for high density devices.
From our point of view it is easier to achieve the same result by changing the shape of the used heat spot to lower the thermal gradient, which also directly influences the ERTW, and thus fundamental AC and DC noise.

\section{Acknowledgements}
The authors would like to thank the Vienna Science and Technology Fund (WWTF) under grant MA14-044 and the Austrian Science Fund (FWF) under grants F4112 SFB ViCoM and I 2214-N20, for financial support. The support from the CD-laboratory AMSEN (financed by the Austrian Federal Ministry of Economy, Family and Youth, the National Foundation for Research, Technology and Development) was acknowledged. 


%

\end{document}